\pdfoutput=1
\documentclass[11pt]{article}

\usepackage[preprint]{acl}

\usepackage{times}
\usepackage{latexsym}
\usepackage{amsmath,amssymb}
\usepackage{geometry}
\usepackage{bm}
\usepackage{mathptmx}
\usepackage{graphicx}
\usepackage{booktabs}
\usepackage{url}
\usepackage[T1]{fontenc}

\usepackage[utf8]{inputenc}

\usepackage{microtype}

\usepackage{inconsolata}

\usepackage{graphicx}

\usepackage{amssymb} 
\usepackage{longtable}
\usepackage{multirow}
\usepackage{csvsimple}
\usepackage{arydshln}
\usepackage{pifont}
\usepackage{xcolor}
\usepackage{enumitem}

\definecolor{darkcheck}{RGB}{0,200,0}    % 更暗的绿色
\definecolor{darkcross}{RGB}{200,0,0}    % 更暗的红色
\newcommand{\cmark}{\textcolor{darkcheck}{\ding{51}}}
\newcommand{\xmark}{\textcolor{darkcross}{\ding{55}}}
\title{Rhythm of Opinion: A Hawkes-Graph Framework for Dynamic Propagation Analysis}

\author{
    \textbf{
        Yulong Li\textsuperscript{1,2*}, 
        Zhixiang Lu\textsuperscript{1*}, 
        Feilong Tang\textsuperscript{2,3}, 
        Simin Lai\textsuperscript{2,3}, 
        Ming Hu\textsuperscript{1}, 
        Yuxuan Zhang\textsuperscript{1},
    }\\
    \textbf{Haochen Xue\textsuperscript{1}, 
        Zhaodong Wu\textsuperscript{2}, 
        Imran Razzak\textsuperscript{2\textdagger}, 
        Qingxia Li\textsuperscript{4\textdagger},
        Jionglong Su\textsuperscript{1\textdagger}}\\
    \hspace*{-2em} 
    \fontsize{9.5pt}{11.6pt}\selectfont{\textsuperscript{1} School of Artificial Intelligence and Advanced Computing, Xi'an Jiaotong-Liverpool University} \\
    \fontsize{9.5pt}{11.6pt}\selectfont{\textsuperscript{2} Mohamed bin Zayed University of Artificial Intelligence} \\
    \fontsize{9.5pt}{11.6pt}\selectfont{\textsuperscript{3} Monash University} \\
    \fontsize{9.5pt}{11.6pt}\selectfont{\textsuperscript{4} Fisk University} \\
    \vspace{-0.1em}
    Imran.Razzak@mbzuai.ac.ae, qli@fisk.edu, Jionglong.Su@xjtlu.edu.cn
}

\begin{document}
\maketitle
\begin{abstract}
The rapid development of social media has significantly reshaped the dynamics of public opinion, resulting in complex interactions that traditional models fail to effectively capture. To address this challenge, we propose an innovative approach that integrates multi-dimensional Hawkes processes with Graph Neural Network, modeling opinion propagation dynamics among nodes in a social network while considering the intricate hierarchical relationships between comments. The extended multi-dimensional Hawkes process captures the hierarchical structure, multi-dimensional interactions, and mutual influences across different topics, forming a complex propagation network. Moreover, recognizing the lack of high-quality datasets capable of comprehensively capturing the evolution of public opinion dynamics, we introduce a new dataset, VISTA. It includes 159 trending topics, corresponding to 47,207 posts, 327,015 second-level comments, and 29,578 third-level comments, covering diverse domains such as politics, entertainment, sports, health, and medicine. The dataset is annotated with detailed sentiment labels across 11 categories and clearly defined hierarchical relationships. When combined with our method, it offers strong interpretability by linking sentiment propagation to the comment hierarchy and temporal evolution. Our approach provides a robust baseline for future research.

\end{abstract}

\section{Introduction}

In the era of rapidly developing social media, the transmission patterns and formation mechanisms of public opinion have undergone profound changes. Public opinion is inherently complex and multi-faceted, driven by individual behavior, social influence, and external stimuli such as news events or policy decisions. Social media platforms have transformed public discussions from unidirectional transmission via traditional media to dynamic inter-

\begin{figure}[h]
  \centering
  \includegraphics[width=1\linewidth]{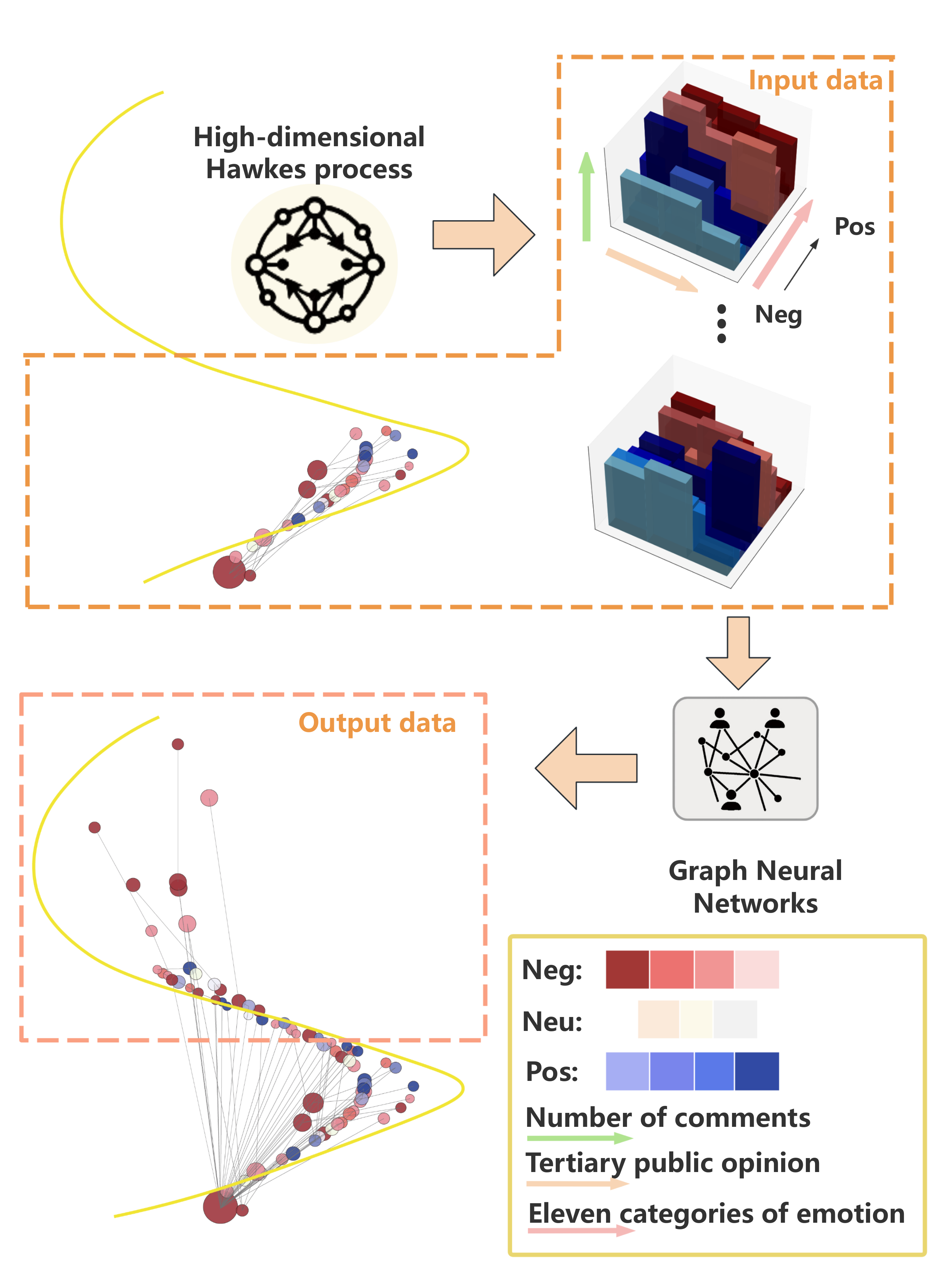}
  \caption{Opinion Dynamics Modeling illustrating two interconnected components: High-dimensional Hawkes Process (top left) and Graph Neural Network-based Sentiment Analysis (middle right).
  \label{fig:fig1} 
    }
 % 控制 caption 与正文的间距
\vspace{-1em}
\end{figure}

%%%%%%%%%%%
\noindent  actions among individuals, groups, and automated systems. This complexity has prompted researchers  to model the dynamic process of opinion dissemination from multiple perspectives. However, existing methods still have many limitations when addressing multi-level public opinion propagation \citep{parsegov2016novel}.\\

%%%%%%%%%%
\vspace{-1em}
Opinion dynamics modeling investigates and predicts how opinions form, spread, and evolve among individuals or groups in social networks using mathematical models, computational methods, and simulation techniques. Current research faces three major challenges. First, time series-based models capture the temporal dynamics of public opinion by forecasting changes over time. However, they overlook the hierarchical structure and complex interactions among comments. Second, network diffusion theory models the propagation paths of public opinion by representing the nodes and edges of social networks. However, most approaches assume static networks, making it difficult to capture temporal propagation dynamics. Third, although agent-based simulation models \citep{gurcan2024llm, muric2022large} can simulate micro-level interaction rules, their low reliance on real-world data results in limited scalability and interpretability in large-scale scenarios. To address these issues, we propose a method that integrates multidimensional Hawkes processes with Graph Neural Networks (GNNs), as illustrated in Figure\ref{fig:fig1}. This approach not only accounts for the temporal dynamics of opinion propagation but also effectively simulates the interactions among nodes in social networks, particularly capturing the complex hierarchical relationships among comments.

%%%%%
Existing publicly available datasets on opinion propagation face significant limitations in content, structure, and temporal dimensions, restricting their application in dynamic opinion modeling. In terms of content, many datasets lack diversity and representativeness. For example, the Twitter dataset \citep{twitter_dataset} only includes first-level comments and retweet information, failing to reflect the complexity of multi-level opinion propagation; the Reddit dataset \citep{hamilton2017inductive} provides multi-level comment information but is mainly concentrated on a few specific topics, making it difficult to cover opinion propagation patterns in different domains and scenarios. In terms of structure, many datasets do not capture the hierarchical relationships in opinion propagation. For instance, the YouTube dataset \citep{kaggle_youtube_comments} only records video comments and likes but lacks descriptions of tree-like hierarchical relationships between comments. Although some datasets such as FakeNewsNet \citep{shu2018fakenewsnet} focus on the authenticity analysis of information dissemination, they fail to model the hierarchical structure and propagation paths of public opinion. In terms of temporal dimensions, most datasets have limited collection time spans, failing to cover the complete lifecycle of public opinion from outbreak to decline. For example, political event datasets \citep{conover2011political} focus only on short-term political event discussions, while COVID-19 datasets \citep{cinelli2020covid} only record early pandemic opinion propagation, failing to capture the long-term dynamic changes of public opinion. Additionally, on a single platform at the same time, multiple hot topics often influence each other, forming complex propagation networks. Interactions among different topics may lead to the migration of user opinions and behaviors across multiple discussions, thereby affecting the transmission patterns and evolution trajectories of public opinion. However, existing datasets generally fail to fully consider this complexity. 

To solve these problems, we propose a brand new dataset that comprehensively covers multi-level comment structures and includes the mutual influence of multiple hot opinion topics. This dataset not only records the lifecycle of topics, from the outbreak to the decline of public opinion, but also considers the interactions and mutual influences among different topics, forming a complex propagation network. Additionally, the hierarchical structure of comments in the dataset adopts a tree-like relationship, which can more realistically reflect the complex interactions in opinion propagation. Based on this dataset, we propose a method combining multidimensional Hawkes processes and GNN. This method can effectively model the temporal dynamics of opinion propagation, the hierarchical relationships among comments, the complex interactions between nodes in social networks, and the interaction processes among different topics.

%%%%%%%%
Our contributions are as follows: 
\vspace{-1.8mm}
\begin{enumerate}[label=\arabic*),itemsep=-2mm]
    \item We propose a novel approach for modeling opinion propagation by integrating multi-dimensional Hawkes processes with GNN. This method effectively captures temporal dynamics, hierarchical structures, and multi-dimensional interactions in social media discussions.
    \item We release the VISTA dataset, a multi-scale public opinion resource that tracks the complete life cycles of 159 viral topics. The dataset includes over 500,000 hierarchically structured comments, annotated with fine-grained sentiment labels across 11 categories.
    \item We establish a highly interpretable model that provides valuable and deep insights into the evolution of opinions in social media discussions, establishing a solid baseline for future studies using the VISTA dataset.
\end{enumerate}

% (1) We propose a novel approach for modeling opinion propagation by integrating multi-dimensional Hawkes processes with GNN. This method effectively captures temporal dynamics, hierarchical structures, and multi-dimensional interactions in social media discussions; (2) We release the VISTA dataset, a multi-scale public opinion resource that tracks the complete life cycles of 159 viral topics. The dataset includes over 500,000 hierarchically structured comments, annotated with fine-grained sentiment labels across 11 categories; and (3) We establish a highly interpretable model that provides valuable and deep insights into the evolution of opinions in social media discussions, establishing a solid baseline for future studies using the VISTA dataset.

%%%
\section{Related Work}
Opinion propagation has evolved significantly since the mid-20th century, initially grounded in sociological theories such as Katz and Lazarsfeld's Two-Step Flow Theory \citep{katz2017personal}, which emphasizes the role of opinion leaders in information dissemination, and Noelle-Neumann's Spiral of Silence Theory \citep{noelle1974spiral}, which accounts for the silence of minority opinions. With the rise of the internet, time series models such as AutoRegressive Integrated Moving Average (ARIMA) have been used to predict popularity trends in opinion events, although they struggle with nonlinear relationships \citep{box2015time,de2016learning,torres2021deep}. The Susceptible-Infected-Recovered (SIR) model \citep{woo2011sir} is adapted for opinion propagation but assumes static networks, limiting its applicability in dynamic contexts \citep{anderson1991infectious,woo2011sir}. Complex network theories, such as Barabási and Albert's scale-free network \citep{barabasi1999emergence} as well as Watts and Strogatz's small-world theory \citep{watts1998collective}, provide further insights into network structures. However, methods such as Graph Convolutional Networks (GCNs) \citep{kipf2016semi} also assume static structures and cannot fully capture evolving dynamics \citep{kipf2016semi,wu2020comprehensive}. Hierarchical models \citep{yang2016hierarchical,jin2014news} address stepwise information spread, but lack temporal dynamics \citep{newman2011structure,arenas2008synchronization}.

From 2015 to 2020, deep learning algorithm such as LSTM and RNN models \citep{wu2020comprehensive} achieve a breakthrough in modeling opinion propagation, capturing long-term dependencies but neglecting opinion structure. GNNs \citep{longa2023graph} aim to model both temporal and structural evolution, but challenges still remain in multi-level propagation \citep{longa2023graph,kazemi2020representation}. Graph Attention Networks (GATs) and other models \citep{rath2021scarlet} have improved predictive performance, with recent research increasingly focused on multi-dimensional aspects, including temporal dynamics, hierarchical structures, and emotional diffusion \citep{chen2023dynamic}. Cross-topic propagation models gain attention during the COVID-19 pandemic, examining the timing and sequencing of social media posts \citep{yin2020quantify}, but unified frameworks integrating these multi-dimensional features are still lacking \citep{parsegov2016novel,cao2019dynamic}. Agent-based simulations, while useful for modeling complex behaviors, face scalability and validation challenges \citep{gurcan2024llm, muric2022large, epstein2012generative}. Despite significant progress, a unified method to simultaneously capture temporal, structural, and multi-dimensional features in opinion propagation remains an open challenge.

\begin{figure}[t]
  \centering
  \includegraphics[width=1\linewidth]{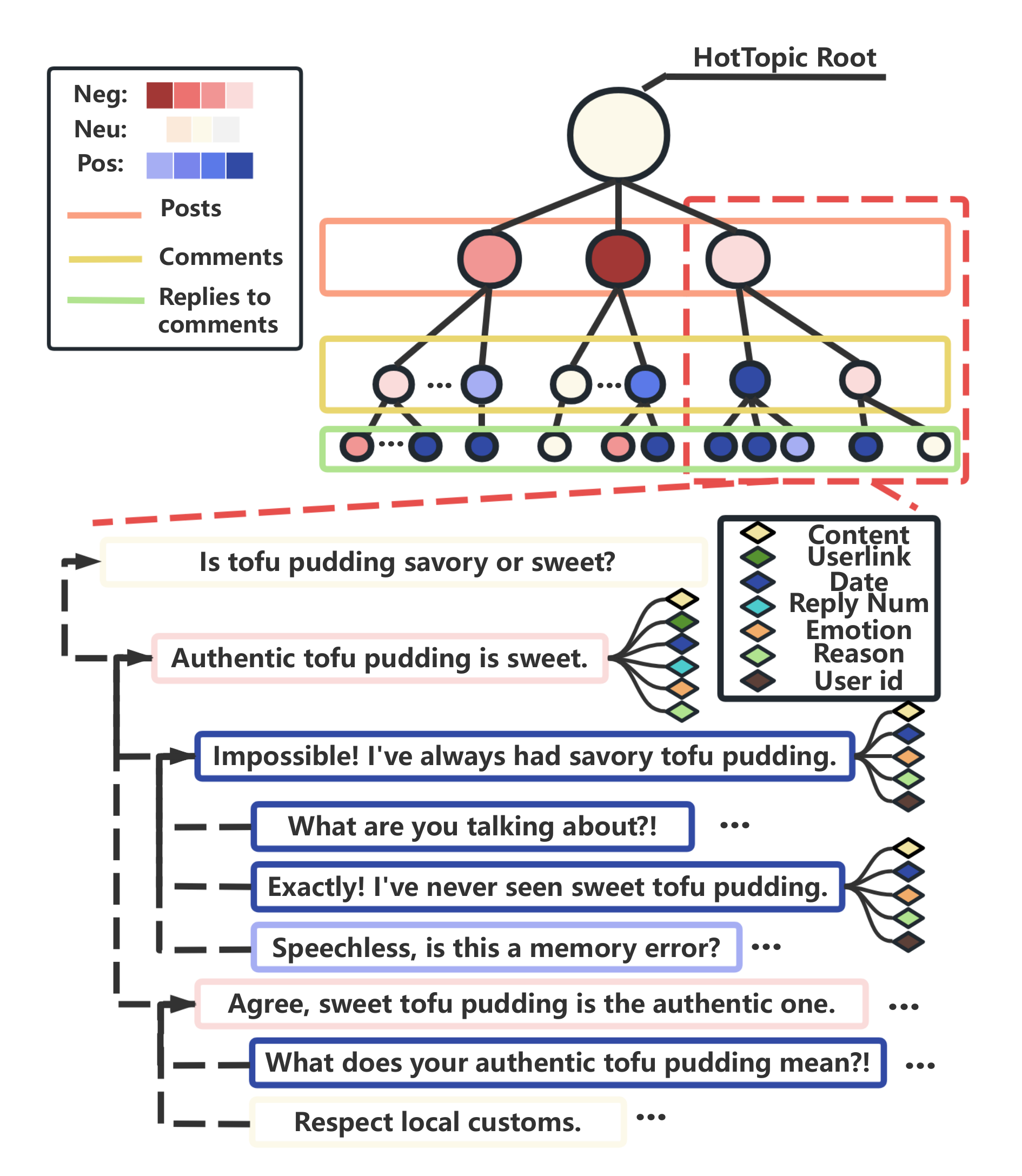}
  \caption{Illustrative Diagram of VISTA Dateset.The replies to comments, comments, and posts together form a complete opinion event in our dataset.
    }
    \label{fig:fig2}
    \vspace{-4mm}
\end{figure}

\section{Dataset Construction}

% To support research on opinion propagation, we obtain comments and interaction records related to trending events from Weibo (\url{https://www.weibo.com}), covering the period from 2024 to early 2025. We monitor newly generated comments in real-time based on specified keywords and event lists to ensure coverage of the complete opinion cycle, from the initial outbreak of events to the decline in attention. After merging the collected raw data, we perform deduplication and format standardization. Subsequently, we employ a custom dictionary and multilingual detection techniques to filter out comments with no substantive content or encoding anomalies. We rigorously screen and remove records containing sensitive personal information. Following this preliminary filtering, we standardize the timestamps of comments to facilitate subsequent temporal analysis and hierarchical structure construction.
To support research on opinion propagation, we collect real-time comments and interaction records from Weibo \citep{weibo} related to trending events, spanning from 2024 to early 2025. Using specified keywords and event lists, we capture the full opinion cycle, from event inception to attention decline. After data merging, we perform deduplication, standardize formats, and apply a custom dictionary and multilingual detection to remove irrelevant or anomalous comments. We also filter out sensitive personal information. Finally, timestamps are standardized for temporal analysis and hierarchical structure construction.

\begin{table*}[h]
  \centering
  
  \resizebox{1\textwidth}{!}{
    \begin{tabular}{cclrrrrrrrrrrrr}
        \midrule
        \multirow{2}[1]{*}{} & \multirow{2}[1]{*}{\textbf{Category}} & \multicolumn{1}{c}{\multirow{2}[1]{*}{\textbf{Type}}} & \multicolumn{12}{c}{\textbf{Emotion classification}} \\
              &       &       & \multicolumn{1}{c}{\textbf{A}} & \multicolumn{1}{c}{\textbf{An}} & \multicolumn{1}{c}{\textbf{S}} & \multicolumn{1}{c}{\textbf{F}} & \multicolumn{1}{c}{\textbf{C}} & \multicolumn{1}{c}{\textbf{N }} & \multicolumn{1}{c}{\textbf{Ca}} & \multicolumn{1}{c}{\textbf{O}} & \multicolumn{1}{c}{\textbf{H}} & \multicolumn{1}{c}{\textbf{Ex}} & \multicolumn{1}{c}{\textbf{El}} & \multicolumn{1}{c}{\textbf{Total}} \\
        \midrule
        \multirow{6}[6]{*}{\textbf{Basic}} & \multirow{2}[2]{*}{\textbf{P}} & \textbf{Normal} & 2529  & 4479  & 3362  & 1866  & 1552  & 3773  & 429   & 11469 & 9784  & 7343  & 621   & 47207 \\
              &       & \textbf{Abnormal} & 0     & 0     & 0     & 0     & 0     & 0     & 0     & 0     & 0     & 0     & 0     & 0 \\
             & \multirow{2}[2]{*}{\textbf{C}} & \textbf{Normal} & 11738 & 10477 & 8398  & 9354  & 5409  & 34412 & 4447  & 74039 & 57331 & 37178 & 12587 & 265370 \\
              &       & \textbf{Abnormal} & 1911  & 1829  & 1006  & 1549  & 1056  & 10873 & 647   & 17443 & 13484 & 9574  & 2273  & 61645 \\
            & \multirow{2}[2]{*}{\textbf{R}} & \textbf{Normal} & 2670  & 2145  & 619   & 1044  & 695   & 10318 & 188   & 2616  & 3341  & 1276  & 45    & 24957 \\
              &       & \textbf{Abnormal} & 343   & 256   & 49    & 150   & 162   & 2024  & 10    & 564   & 721   & 305   & 37    & 4621 \\
        \hdashline
        \multirow{4}[4]{*}{\textbf{Social}} & \multirow{2}[2]{*}{\textbf{CL}} & \textbf{Normal} & 407   & 468   & 311   & 533   & 115   & 1038  & 96    & 2856  & 1265  & 1227  & 161   & 8477 \\
              &       & \textbf{Abnormal} & 50    & 121   & 60    & 64    & 44    & 236   & 17    & 856   & 495   & 365   & 74    & 2382 \\
             & \multirow{2}[2]{*}{\textbf{RL}} & \textbf{Normal} & 1680  & 1095  & 287   & 641   & 357   & 3730  & 70    & 964   & 1171  & 402   & 16    & 10413 \\
              &       & \textbf{Abnormal} & 231   & 111   & 25    & 94    & 38    & 710   & 2     & 160   & 241   & 84    & 12    & 1708 \\
        \bottomrule

    \end{tabular}%
    }
  \begin{minipage}[t]{1\textwidth}
    \raggedright
    \small *The dataset used in this study does not contain any fake accounts (or astroturfing)\\
     *P:posts. C:comments. R:replies. CL:comments with likes. RL:replies with likes.\\
     *A:Angry, An:Anxious, S:Sad, F:Frustrated, C:Consoling, N:Neutral, Ca:Calm, O:Optimistic, H:Happy, Ex:Excited, El:Elated\\
  \end{minipage}
  \caption{Statistics of VISTA Dataset}
  \label{tab:dataset_analysis}%
  \vspace{-3mm}
\end{table*}%

% While retaining the content of the comments, we record and correct the parent comment IDs for each comment to reconstruct the tree-like topological structure of comments and replies. To achieve this, we utilize hash mapping and indexing to match each comment with its parent comment, tracing back layer by layer to the deepest child comment and limiting the comment depth to a maximum of three levels. A three-level structure is chosen to capture people's opinions on the post, their comments, and the supporting or opposing voices in the discussion, ensuring a comprehensive yet efficient representation of opinion propagation. In the rare cases where parent IDs are missing or point to deleted comments, we treat these comments as top-level comments and mark them separately. Upon completion of this process, the comment data manifests as multiple directed trees, with each comment possessing an accurate hierarchical label. As shown in Figure \ref{fig:fig2}, the VISTA dataset includes the structure of opinion events, where replies to comments, comments, and posts together form a complete opinion event. The statistical breakdown of the VISTA dataset is provided in Table \ref{tab:dataset_analysis}, where we categorize the dataset by different levels of comments and emotional annotations.
To retain the content of comments, we reconstruct the tree-like structure of comments and replies by recording and correcting the parent comment IDs. Using hash mapping and indexing, we match each comment with its parent, tracing back layer by layer to the deepest child comment while limiting the depth to a maximum of three levels. This three-level structure captures opinions on posts, comments, and supporting or opposing voices, providing an efficient and comprehensive representation of opinion propagation. In cases where parent IDs are missing or refer to deleted comments, we treat them as top-level comments and mark them separately. After this process, the comment data forms directed trees, with each comment assigned an accurate hierarchical label. As illustrated in Figure \ref{fig:fig2}, the VISTA dataset captures the full structure of opinion events, including replies, comments, and posts. The dataset's statistical breakdown, categorized by comment levels and emotional annotations, is provided in Table \ref{tab:dataset_analysis}.

\begin{table}
  \centering
  \resizebox{0.5\textwidth}{!}{
  \begin{tabular}{l| l l l l l l}
    \hline
    \textbf{Dataset} & \textbf{Md}  & \textbf{Cd} & \textbf{Lt} & \textbf{Cpoe} & \textbf{El}  \\
    \hline
    Twitter Dataset & \cmark & \cmark  & \xmark & \xmark & \cmark \text{*} \\
    Reddit Dataset & \cmark & \cmark  & \xmark & \xmark & \xmark \\
    YouTube Dataset & \cmark & \cmark  & \xmark & \xmark & \xmark \\
    FakeNewsNet Dataset & \xmark & \xmark  & \cmark & \xmark & \cmark \\
    Political Event Dataset & \xmark & \xmark  & \xmark & \xmark & \xmark \\
    COVID-19 Dataset & \xmark & \xmark  & \cmark & \xmark & \xmark \\
    \textbf{Ours} & \cmark & \cmark & \cmark& \cmark & \cmark  \\
    \hline
  \end{tabular}
  }
  \\

    \noindent
    \hspace{1.8mm}
    \begin{minipage}[t]{0.48\textwidth}
        \raggedright
        \small
             Md: Multilevel data.
             Cd: Content diversity.
             Lt: Length of time. 
             Cpoe: Complete public opinion event. 
             El: Emotional label. 
        \textbf{*}:Partial annotation
    \end{minipage}
  \caption{Multidimensional comparison of datasets}
  \label{tab:datasets_comparison}
  \vspace{-3mm}
\end{table}
%%%
% For sentiment and event label generation, we use the GLM-4-plus model for automated annotation of comments. The results in 11 levels of sentiment labels subdivided into three major categories: negative, neutral, and positive, along with the model's rationale for sentiment judgments. To enhance the reliability of annotations, we first manually annotate a subset of comments with high precision. We use these high-quality samples for prompt engineering and fine-tuning of the large model before performing batch inference on the remaining comments. During the secondary review process, if any significant discrepancies with manual judgments are detected, we modify or revert the automatic annotations accordingly. Finally, we conduct spot checks to verify the overall quality of the dataset, ensuring that the mean Cohen’s Kappa of the ten final spot checks for sentiment annotation reaches 0.85 . We compare the multidimensional aspects of various opinion propagation datasets in Table \ref{tab:datasets_comparison}, highlighting the strengths and limitations of each dataset. Additionally, we assign corresponding event labels to each comment, in conjunction with recorded user attributes, ensuring the traceability of each comment within the social network and the context of public opinion.
For sentiment and event label generation, we use the GLM-4-plus model for automated comment annotation, yielding 11 sentiment levels categorized into negative, neutral, and positive, along with the model's rationale. To ensure annotation reliability, we first manually label a subset of high-quality comments, using them for prompt engineering and fine-tuning before applying batch inference to the remaining comments. During a secondary review, significant discrepancies with manual annotations are corrected. Finally, spot checks are performed to verify dataset quality, achieving a mean Cohen’s Kappa of 0.85 for sentiment annotation across ten final checks. We compare the multidimensional aspects of various opinion propagation datasets in Table \ref{tab:datasets_comparison}, highlighting their strengths and limitations. Event labels are also assigned to each comment, along with user attributes, ensuring traceability within the social network and public opinion context.

% \noindent 
\section{Task Definition}
Suppose there are $M$ parallel ``hot topics'' in a given time period, each represented by a root post $H_m$ for $m \in \{1,\dots,M\}$. Under these root posts, users continuously produce hierarchical comments: if a comment replies directly to a hot topic, it is considered level 1; if it replies to a comment at the previous level, it spawns a new node in that thread, increasing the level by 1.  Formally, we denote each observed comment event as 
\[
\mathcal{E} \;=\; \bigl\{\, e_i \;\bigm|\; e_i=(h_i,\,l_i,\,p_i,\,t_i)\bigr\},
\]
where $h_i \in \{1,\dots,M\}$ indicates the hot topic to which the comment belongs ($i$ represents an individual comment event, which includes its topic, level, parent comment, and timestamp),  $l_i \ge 1$ is the comment's level within that topic, $p_i$ is the parent comment or root post, and $t_i$ is the posting timestamp.

Because multiple hot topics may exhibit mutual influence (for example, content overlap or shared user bases across different topics), comment evolution can manifest both \emph{cross-topic} mutual excitation and \emph{hierarchical expansion} within each individual chain. Therefore, given a historical comment event set $\mathcal{E}_{\le T}$ (i.e., all comments observed up to some time $T$), our goal is to learn a predictive function
\[
F\bigl(\mathcal{E}_{\le T}\bigr) \;\mapsto\; \widehat{\mathcal{E}}_{>T},
\]
where $\mathcal{E}_{\le T}$ denotes all observed comments at or before $T$, and $\widehat{\mathcal{E}}_{>T}$ denotes the predicted comments that may appear during the future time interval $(T,\, T+\Delta]$. 

Concretely, the prediction covers two complementary dimensions: 
\vspace{-0.5em}
\begin{itemize}
    \item  \textbf{Temporal dimension}, it focuses on predicting when new comments will emerge and how the volume of comments will change over time across various topics and hierarchical levels.
    \vspace{-0.5em}
    \item  \textbf{Structural dimension}, it focuses on how newly added comments connect to parent nodes, potentially spanning diverse topics or levels, thus driving the dynamic evolution of the hierarchical structure.
\end{itemize}
\vspace{-0.25em}

The core of the task lies in simultaneously accounting for cascading responses within the hierarchical structure (where higher-level comments trigger replies at subsequent levels) and sentiment propagation patterns to provide a comprehensive prediction of future comment behavior. This requires not only modeling the dynamic temporal and structural evolution of the discussion network but also effectively capturing how sentiment propagates through comments, enabling more accurate forecasting of the evolution of public opinion on social media platforms.

\section{Methodology}
% \newpage
\begin{figure*}[t]
  \centering
  \includegraphics[width=1.1\linewidth]{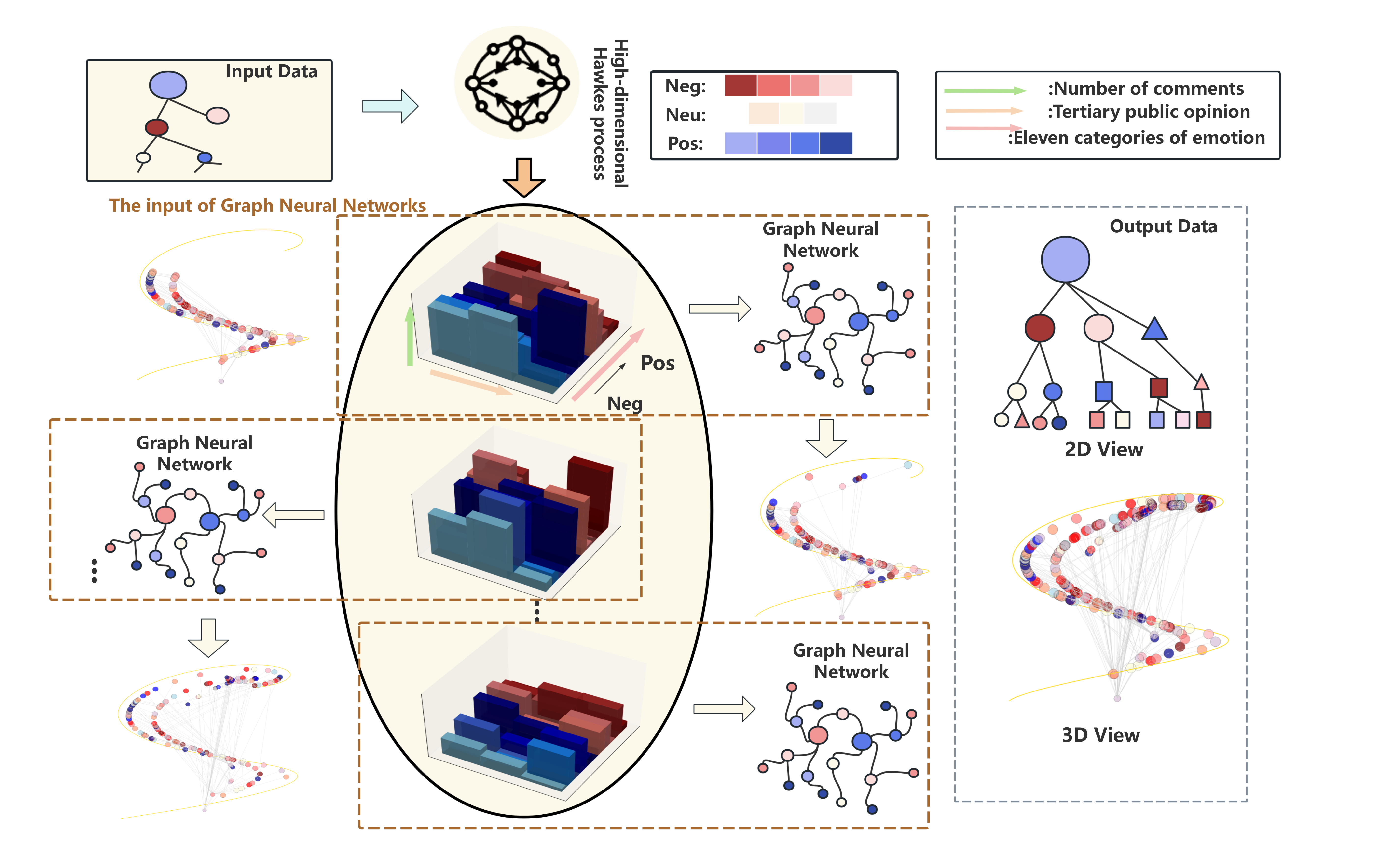}
  \caption{Architecture of the Social Network Sentiment Analysis Model. Our framework analyzes sentiment in social networks using a high-dimensional Hawkes process and graph neural networks. The Hawkes process predicts comment volume and sentiment, while GNN learns the social structure, classifies nodes, and constructs a predicted network. Through iterations, the framework achieves sentiment analysis.}
  \label{fig:fig3}
  \vspace{-3mm}
\end{figure*}

As illustrated in Figure \ref{fig:fig3}, the architecture of our Social Network Sentiment Analysis model combines a high-dimensional Hawkes process with a graph neural network to model sentiment dynamics.

%%%
Our method integrates multi-level comments with fine-grained sentiment levels. When incorporating the Hawkes process into sentiment modeling, the main challenge lies in obtaining direct predictions for the number of comments associated with different sentiment opinions. To address this, we model the Hawkes process in higher dimensions by combining hierarchy and sentiment levels as two discrete attributes. Each dimension corresponds to a specific ``(hierarchy, sentiment level)'' pair, enabling the intensity function to estimate the number of comments associated with various sentiment opinions. The high-dimensional Hawkes process captures the temporal evolution of comments under different sentiment labels and hierarchy levels. The output is then used to construct a GNN, further capturing the dynamic propagation structure and sentiment diffusion patterns among comments.

Let \( l \) denote the hierarchy level and \( c \) denote the sentiment level, where there are \( L \) hierarchy levels and \( C \) sentiment levels. In this paper, \( L = 3 \) and \( C = 11 \). Each comment is mapped to a dimension  \( \omega = (l, c) \), where \( 1 \leq l \leq L \) and \( 1 \leq c \leq C \). The Cartesian product of hierarchy levels and sentiment levels forms a new set of dimensions:
\[
\Omega = \left\{ (l, c) \,\middle\vert\, l \in \{1,\dots,L\},\, c \in \{1,\dots,C\} \right\},
\]
where \( \Omega \) represents all possible combinations of hierarchy levels and sentiment levels, and \( \omega \in \Omega \) corresponds to a specific type of event. If a comment occurs at time \( t_i^{\omega} \), then it is assigned to dimension \( \omega \). Based on this, the intensity function of the high-dimensional Hawkes process is defined as:
\begin{equation}
\label{eq:hawkes_intensity}
\lambda_{\omega}(t) = \mu_{\omega} + \sum_{\omega' \in \Omega} \sum_{t_j^{\omega'} \leq t} \alpha_{\omega,\omega'} \phi_{\omega,\omega'}(t - t_j^{\omega'}),
\end{equation}
where \( \mu_{\omega} \) denotes the baseline arrival rate for dimension \( \omega \), \( \alpha_{\omega,\omega'} \) measures the excitation strength from dimension \( \omega' \) to dimension \( \omega \), and \( \phi_{\omega,\omega'}(\tau) \) is the decay kernel function. In this case, the decay kernel function is defined as an exponential kernel, \( \phi_{\omega,\omega'}(\tau) = e^{-\beta_{\omega,\omega'}\tau} \mathbf{1}_{\{\tau \geq 0\}} \), where \( \beta_{\omega,\omega'} \) is the decay coefficient and \(\mathbf{1}_{\{\tau \geq 0\}}\) is indicator function.

To estimate \( \mu_{\omega} \), \( \alpha_{\omega,\omega'} \), and \( \beta_{\omega,\omega'} \), we employ the maximum likelihood method. Given an observation window \( [0, T] \), let \( \{t_1^{\omega}, t_2^{\omega}, \dots, t_{N_{\omega}}^{\omega}\} \) denote the occurrence times of all events (comments) in dimension \( \omega \), with \( N_{\omega} \) being the total number of events. The log-likelihood function of the high-dimensional Hawkes process is expressed as:
\begin{equation}
\label{eq:hawkes_likelihood}
\mathcal{L} = \sum_{\omega \in \Omega} \Bigl[ \sum_{i=1}^{N_{\omega}}\ln\bigl(\lambda_{\omega}(t_i^{\omega})\bigr) - \int_{0}^{T} \lambda_{\omega}(t) \, dt \Bigr],
\end{equation}
where \( T \) is the end of the observation window.

\noindent \textbf{Log-likelihood and Gradient Derivations}

The parameters to be learned are:
\begin{align}
\boldsymbol{\theta} = \bigl\{\, & \{\mu_{\omega}\}_{\omega \in \Omega},\; 
\{\alpha_{\omega,\omega'}\}_{\omega, \omega' \in \Omega},\; 
\{\beta_{\omega,\omega'}\}_{\omega, \omega' \in \Omega} \,\bigr\}.
\end{align}

%%%%
The gradient of \( \mathcal{L} \) taken with respect to \( \mu_{\omega} \) is:
\begin{align}
\frac{\partial \mathcal{L}}{\partial \mu_{\omega}} = \sum_{i=1}^{N_{\omega}} \frac{1}{\lambda_{\omega}(t_i^{\omega})} - \int_{0}^{T} 1 \, dt = \sum_{i=1}^{N_{\omega}} \frac{1}{\lambda_{\omega}(t_i^{\omega})} - T.
\end{align}

% For the gradient with respect to \( \alpha_{\omega,\omega'} \), it includes the log term and the integral term. The gradient of the log term can be calculated as discussed in Appendix Section \ref{appendix:A}.The gradient of the log term can be calculated as:

%%%%
For the gradient with respect to \( \alpha_{\omega,\omega'} \), it includes the log term and the integral term. The gradient of the log term can be calculated as:\\

\begin{align}
\sum_{i=1}^{N_{\omega}} \ln\!\bigl(\lambda_{\omega}(t_i^{\omega})\bigr) \rightarrow \sum_{i=1}^{N_{\omega}} \frac{1}{\lambda_{\omega}(t_i^{\omega})} \cdot \frac{\partial}{\partial \alpha_{\omega,\omega'}} \lambda_{\omega}\bigl(t_i^{\omega}\bigr).
\end{align}
For a detailed mathematical proof, refer to Appendix Section \ref{appendix:A}.

Using Equation \eqref{eq:hawkes_intensity}, we derive:
\begin{align}
\frac{\partial}{\partial \alpha_{\omega,\omega'}} \lambda_{\omega}(t_i^{\omega}) = \sum_{j=1}^{N_{\omega'}} e^{-\beta_{\omega,\omega'} \left(t_i^{\omega} - t_j^{\omega'}\right)} \mathbf{1}_{\{t_j^{\omega'} < t_i^{\omega}\}}.
\end{align}
where \( N_{\omega'} \) is the total number of events of type \( \omega' \) in the observed time window \([0, T]\). 

The integral term can be rewritten as:
\begin{multline}
\int_{0}^{T} \partial_{\alpha} \lambda_{\omega}(t) \, dt 
= \sum_{j=1}^{N_{\omega'}} \int_{t_j^{\omega'}}^{T} \partial_{\alpha} \left( \alpha_{\omega,\omega'} 
e^{-\beta_{\omega,\omega'} (t - t_j^{\omega'})} \right) dt \\
= \sum_{j=1}^{N_{\omega'}} \int_{t_j^{\omega'}}^{T} e^{-\beta_{\omega,\omega'} (t - t_j^{\omega'})} \, dt \\
= \sum_{j=1}^{N_{\omega'}} \frac{1}{\beta_{\omega,\omega'}} \left[ 1 - e^{-\beta_{\omega,\omega'} (T - t_j^{\omega'})} \right].
\end{multline}
where  \(\partial_{\alpha} \equiv \frac{\partial}{\partial \alpha_{\omega,\omega'}}\).

% Combining the terms, we can arrive at the gradient with respect to \( \alpha_{\omega,\omega'} \) as 

%%%
Combining the terms, we can arrive at the gradient with respect to \( \alpha_{\omega,\omega'} \) as:\\

\begin{align}
\frac{\partial \mathcal{L}}{\partial \alpha_{\omega,\omega'}} = \sum_{i=1}^{N_{\omega}} \frac{1}{\lambda_{\omega}(t_i^{\omega})} \sum_{j=1}^{N_{\omega'}} e^{-\beta_{\omega,\omega'} \left(t_i^{\omega} - t_j^{\omega'}\right)} \mathbf{1}_{\{t_j^{\omega'} < t_i^{\omega}\} }\notag \\
- \sum_{j=1}^{N_{\omega'}} \frac{1}{\beta_{\omega,\omega'}} \left[1 - e^{-\beta_{\omega,\omega'} (T - t_j^{\omega'})} \right].
\end{align}
For a detailed mathematical proof, refer to Appendix Section \ref{appendix:B}.
The gradient with respect to \( \beta_{\omega,\omega'} \) follows a similar derivation. For a detailed mathematical proof, refer to Appendix Section \ref{appendix:C}.
 For the Hawkes process to remain stable and avoid exponential growth, the following stability condition is required for each dimension \( \omega \):
\begin{align}
\sum_{\omega' \in \Omega} \frac{\alpha_{\omega,\omega'}}{\beta_{\omega,\omega'}} < 1.
\end{align}
Otherwise, it leads to a critical or supercritical state, resulting in exponential comment growth and loss of stationarity.

\noindent \textbf{Prediction of Comment Counts}

To predict the number of comments for a specific hierarchy-sentiment pair within the interval \( [T, T+\Delta] \), we integrate the intensity function over the interval. For any \( \omega = (l, c) \), the expected number of events occurring within \( [T, T+\Delta] \) is given by:
\begin{align}
{
\hat{N}_{\omega}(T, T+\Delta) = \int_{T}^{T+\Delta} \lambda_{\omega}(t) \, dt.
}
\end{align}
For scenarios focusing on all sentiment levels within a specific hierarchy level or all hierarchy levels within a specific sentiment level, the predictions for the corresponding dimensions \( \omega \) are aggregated accordingly. The high-dimensional Hawkes model is trained using stochastic gradient methods, enabling the simultaneous estimation of all parameters \( \mu_{\omega}, \alpha_{\omega,\omega'} \), and \( \beta_{\omega,\omega'} \). This framework facilitates both independent and combined calculations of intensities across dimensions during inference or prediction.

\noindent\textbf{Graph Neural Network Modeling}

The GNN models the evolution of opinion propagation. Its primary tasks are node classification and edge prediction. Node classification aims to predict the sentiment level of each comment node, while edge prediction seeks to infer the propagation paths and their strengths between comments. Based on the predicted comments and their sentiment distributions generated by the high-dimensional Hawkes process, we construct an opinion propagation graph, denoted as \( \mathcal{G} = (\mathcal{V}, \mathcal{E}) \), where \( \mathcal{V} \) represents the set of comment nodes, and \( \mathcal{E} \) represents the parent-child relationships between comments.

Each node's features are composed of the intensity function values \( \lambda_{\omega}(t) \) generated by the Hawkes process and the sentiment distribution \( q_c(v) \). Here, \( \lambda_{\omega}(t) \) represents the arrival intensity of the comment, and \( q_c(v) \) denotes the probability that node \( v \) belongs to sentiment level \( c \in \{1, 2, \dots, C\} \). Specifically, the sentiment distribution is defined as:
\begin{align}
q_c(v) = \frac{\lambda_{(l,c)}(t)}{\sum_{c'=1}^{C} \lambda_{(l,c')}(t)},
\end{align}
where \( \lambda_{(l,c)}(t) \) is the intensity function value at time \( t \) for hierarchy level \( l \) and sentiment level \( c \), and \( C \) is the total number of sentiment categories. This distribution captures the sentiment inclination of the node. The edge features \( \mathbf{e}_{uv} \) include the time difference between nodes, \( \Delta t = t_v - t_u \), \( \Delta t > 0 \), and the excitation strength \( \alpha_{\omega, \omega'} \). The former measures the time interval between comments, while the latter reflects the propagation relationships between different sentiments and hierarchy levels.

In the graph, node embeddings are updated using a message-passing mechanism. Let \( t \) denote the current time slice and the embedding of node \( v \) be defined as \( \mathbf{h}_v^{(t)} \). The updated rule for node embeddings is:
\begin{align}
\mathbf{h}_v^{(t+1)} = \sigma \left( \mathbf{W}_1 \mathbf{h}_v^{(t)} + \sum_{u \in \mathcal{N}(v)} \mathbf{W}_2 \mathbf{e}_{uv} \mathbf{h}_u^{(t)} \right),
\end{align}
where \( \mathbf{h}_v^{(t+1)} \) is the embedding of node \( v \) at time \( t+1 \), \( \mathcal{N}(v) \) is the set of neighbors of node \( v \), \( \mathbf{e}_{uv} \) is the feature vector of edge \( (u, v) \), \( \mathbf{W}_1, \mathbf{W}_2 \) are learnable linear transformation matrices, and \( \sigma(\cdot) \) is a non-linear activation function. This update formula aggregates features dynamically by combining historical embeddings of the node itself with embeddings of its neighbors, weighted by the edge features.

Based on the updated node embeddings, we classify the sentiment level of each node using the following formula:
\begin{align}
P(c \mid v) = \mathrm{softmax}(\mathbf{W}_c \mathbf{h}_v^{(t+1)}),
\end{align}
where \( \mathbf{W}_c \) is the linear transformation matrix for sentiment classification, and \( P(c \mid v) \) represents the predicted probability that node \( v \) belongs to sentiment level \( c \).

To optimize the accuracy of node classification and edge prediction, we introduce two loss functions: the sentiment prediction loss \( \mathcal{L}_{\text{sentiment}} \) and the graph structure loss \( \mathcal{L}_{\text{struct}} \). The sentiment prediction loss measures the accuracy of the predicted sentiment distribution for nodes and is defined as:
\begin{align}
\mathcal{L}_{\text{sentiment}} = - \sum_{v \in \mathcal{V}_{\text{pred}}} \sum_{c=1}^{C} q_c(v) \cdot \ln P(c \mid v),
\end{align}
where \( \mathcal{V}_{\text{pred}} \) denotes the set of nodes predicted by the Hawkes process, and \( q_c(v) \) is the normalized sentiment distribution output from the Hawkes process, representing the true sentiment label of node \( v \).

The graph structure loss measures the structural differences between the predicted graph \( \mathcal{G}_{\text{pred}} \) and the true graph \( \mathcal{G}_{\text{true}} \). It is defined as:
\begin{align}
\mathcal{L}_{\text{struct}} = \sum_{(u,v) \in \mathcal{E}} \left| e_{uv}^{\text{pred}} - e_{uv}^{\text{true}} \right|,
\end{align}
where \( \mathcal{E} \) is the set of edges in the graph, and \( e_{uv}^{\text{pred}} \) and \( e_{uv}^{\text{true}} \) represent the feature values of edge \( (u, v) \) in the predicted graph and the true graph, respectively, reflecting the differences in propagation paths.

The final optimization objective combines the two losses, and the composite loss function is defined as:
\begin{align}
\mathcal{L}_{\text{total}}  = \lambda_1 \mathcal{L}_{\text{sentiment}} + \lambda_2 \mathcal{L}_{\text{struct}},
\end{align}
where \( \lambda_1 \) and \( \lambda_2 \) are hyperparameters that control the weights of the sentiment loss and the structure loss, respectively. By minimizing \( \mathcal{L}_{\text{total}} \), the graph neural network optimizes the performance of both node classification and edge prediction, thereby capturing the dynamic evolution of opinion propagation in terms of both temporal and structural dimensions.

\section{Evaluation Metrics}

\textbf{Sentiment Prediction Accuracy (SA)}:  
   The sentiment prediction accuracy measures how well the model predicts the sentiment label for each comment. Given the true sentiment label \( c(v) \) and the predicted sentiment label \( \hat{c}(v) \) for each comment node \( v \), the sentiment prediction accuracy \( \text{Acc}_{\text{sentiment}} \) is defined as:
   \[
   \text{Acc}_{\text{sentiment}} = \frac{1}{N} \sum_{v \in \mathcal{V}} \mathbf{1}(\hat{c}(v) = c(v))
   \]
   where \( \mathbf{1}(\cdot) \) is an indicator function that equals 1 when the predicted sentiment matches the true sentiment, and 0 otherwise.

\textbf{Structural Consistency Prediction Accuracy (SCA)}:  
   The structural consistency prediction accuracy measures how well the model predicts the hierarchical structure of comments. Specifically, it checks whether the set of child nodes for each parent node is correctly predicted. The structural consistency prediction accuracy \( \text{Acc}_{\text{structure}} \) is defined as:
   \[
   \text{Acc}_{\text{structure}} = \frac{1}{|\mathcal{V}_{\text{true}}|} \sum_{v \in \mathcal{V}_{\text{true}}} \mathbf{1} \left( \mathcal{C}_v^{\text{pred}} = \mathcal{C}_v^{\text{true}} \right)
   \]
   where \( \mathcal{C}_v^{\text{true}} \) and \( \mathcal{C}_v^{\text{pred}} \) are the sets of child nodes of the parent node \( v \) in the true and predicted graphs, respectively. 
\section{Experiments}

%%%%%
\subsection{Dataset}
To ensure the reliability and fairness of the experimental results, we divide the VISTA dataset into training, validation, and test sets. Specifically, 127 trending topics are used for training the model, covering all topics and levels of comments to ensure the model can learn diverse opinion chain features. Both the validation and test sets consist of 16 trending topics each.

\subsection{Baseline for Dynamic Propagation}
%%%%
Table 3 gives the performance evaluation of the model on the future opinion chain prediction task at three data proportions (15\%, 20\%, and 25\%). As the data proportion increases, both Sentiment Prediction Accuracy and Structural Consistency Prediction Accuracy performance consistently improve, indicating that a larger dataset effectively enhances the model's predictive capabilities. This table provides the baseline performance of our model on the VISTA dataset, confirming the significant impact of data volume on prediction accuracy.

\begin{table}[htbp]
  \centering
  \resizebox{0.5\textwidth}{!}{
    \begin{tabular}{c c c}
        \hline
        % \rowcolor{gray!20} % Light gray background for the header
        \textbf{Data Proportion} & \textbf{SA (Val/Test \%)} & \textbf{SCA (Val/Test \%)} \\
        \hline
        15\% & 19.75/18.31 & 23.41/21.22 \\
        % \hline
        20\% & 24.12/22.19 & 29.34/26.98 \\
        % \hline
        25\% & 29.31/26.99 & 37.29/35.76 \\
        \hline
    \end{tabular}
    }
  \caption{Performance of Future Opinion Chain Prediction with Different Data Proportions. This table presents the evaluation of the model's performance on future opinion chain prediction tasks, based on three different data proportions: 15\%, 20\%, and 25\%. }
  \label{tab:data_proportion_performance}
  \vspace{-4mm}
\end{table}

\section{Conclusion}

In conclusion, we propose a novel and highly interpretable method that combines high-dimensional Hawkes processes with Graph Neural Networks for modeling opinion propagation on social media. By introducing the VISTA dataset, which includes multi-level comment structures, sentiment annotations, and interactions between trending topics, we provide valuable research data for future studies on opinion dynamics. Our method effectively captures the temporal evolution, structural changes, and sentiment diffusion patterns in discussions. In the future, we plan to apply more advanced methods on this dataset to further explore the underlying mechanisms of opinion dynamics.

\section*{Limitations}
Although this study provides valuable insights, there are still some limitations. Our data is limited to Weibo, which may restrict the generalizability of the conclusions, as user behavior may differ across various social platforms and cultural contexts. Additionally, we assume that the data points are independent; however, in practice, interactions within Weibo's fan communities and the self-organized moderation of comments may lead to data correlations, impacting the accuracy of the model. We also assume that the data collection process is noise-free, but in reality, noise and manipulation could affect the reliability of the results. While we have adopted an interpretable model to ensure transparency, this may limit the model's predictive performance, especially when dealing with more complex data. The model may be overly sensitive to Weibo-specific data characteristics, resulting in limited generalizability. We have decided to make all our code publicly available for other researchers to validate and reproduce our experimental results. Future research will consider validating our model on more platforms and datasets to further improve its robustness and generalizability.

\label{sec:bibtex}

\bibliography{custom}

\newpage
\section{Appendix}
\subsection{Ethics Statements }
The dataset used in our study is collected in compliance with ethical research guidelines, ensuring that user privacy and data security are adequately protected. The data primarily consists of social media discussions, including posts, comments, and replies, annotated with multiple emotional categories. To minimize risks associated with sensitive content, we implement a multi-stage filtering process to remove personally identifiable information, explicit material, and offensive language. However, due to the inherent limitations of automated and manual inspection, some residual content may persist, making complete elimination a challenging task. 

Since the dataset originates from real-world social interactions, it may include a small amount of misinformation or subjective opinions, which could influence analytical outcomes and model performance. We release this dataset exclusively for research purposes, aiming to support studies in opinion dynamics, sentiment analysis, and social network modeling. Researchers using this dataset should exercise caution and adhere to ethical guidelines when interpreting the results. 

Our goal is to contribute to the advancement of computational social science and machine learning while ensuring responsible data usage. In future updates, we will continue to expand and refine the dataset, improving both its coverage and filtering mechanisms to enhance its reliability and applicability in diverse research scenarios.

\subsection{A}\label{appendix:A}
Given the high-dimensional Hawkes process intensity function \( \lambda_{\omega}(t) \), the log-likelihood function is defined as:
\begin{align}
\mathcal{L} = \sum_{\omega \in \Omega} \left[ \sum_{i=1}^{N_{\omega}} \ln \left( \lambda_{\omega}(t_i^{\omega}) \right) - \int_0^T \lambda_{\omega}(t) \, dt \right],
\end{align}
where \( t_i^{\omega} \) is the \( i \)-th event time in the \( \omega \)-th dimension. 

The intensity function \( \lambda_{\omega}(t) \) is composed of two terms: the first term \( \sum_{i=1}^{N_{\omega}} \ln \left( \lambda_{\omega}(t_i^{\omega}) \right) \) is the log-likelihood of the observed event times \( t_i^{\omega} \), and the second term \( \int_0^T \lambda_{\omega}(t) \, dt \) is the integral of the intensity function \( \lambda_{\omega}(t) \) over the time interval \( [0, T] \), representing the expected number of events.

The intensity function for the high-dimensional Hawkes process is defined as:
\begin{align}
\lambda_{\omega}(t) = \mu_{\omega} + \sum_{\omega' \in \Omega} \sum_{t_j^{\omega'} \leq t} \alpha_{\omega, \omega'} \phi_{\omega, \omega'}(t - t_j^{\omega'}),
\end{align}
where \( \mu_{\omega} \) is the baseline intensity, representing the event rate in the absence of any historical events, \( \alpha_{\omega, \omega'} \) is the excitation strength from event type \( \omega' \) to \( \omega \), and \( \phi_{\omega, \omega'}(t - t_j^{\omega'}) \) is the decay kernel, typically in the form of an exponential decay.

We need to compute the gradient of the log-likelihood function with respect to \( \mu_{\omega} \):
\begin{align}
\frac{\partial \mathcal{L}}{\partial \mu_{\omega}}.
\end{align}
The gradient with respect to \( \mu_{\omega} \) is derived in two parts.
First, we consider the left part:
\begin{align}
\sum_{i=1}^{N_{\omega}} \ln \left( \lambda_{\omega}(t_i^{\omega}) \right).
\end{align}
By taking the partial derivative with respect to \( \mu_{\omega} \) using the chain rule, we obtain that:
\begin{align}
\frac{\partial}{\partial \mu_{\omega}} \ln \left( \lambda_{\omega}(t_i^{\omega}) \right) = \frac{1}{\lambda_{\omega}(t_i^{\omega})} \cdot \frac{\partial \lambda_{\omega}(t_i^{\omega})}{\partial \mu_{\omega}}.
\end{align}
Since \( \lambda_{\omega}(t) \) only involves \( \mu_{\omega} \) in the first term (i.e., \( \mu_{\omega} \) only affects the baseline part of the intensity function), it yields that
\begin{align}
\frac{\partial \lambda_{\omega}(t)}{\partial \mu_{\omega}} = 1.
\end{align}
Therefore, for each event \( t_i^{\omega} \), we conclude:
\begin{align}
\frac{\partial}{\partial \mu_{\omega}} \ln \left( \lambda_{\omega}(t_i^{\omega}) \right) = \frac{1}{\lambda_{\omega}(t_i^{\omega})}.
\end{align}
Summing over all \( i \), we get:
\begin{align}
\sum_{i=1}^{N_{\omega}} \frac{1}{\lambda_{\omega}(t_i^{\omega})}.
\end{align}
Next, we will tackle the second part:
\begin{align}
- \int_0^T \lambda_{\omega}(t) \, dt.
\end{align}
We note that the above term involves the integral of \( \lambda_{\omega}(t) \) which is a function of \( \mu_{\omega} \). Taking the partial derivative with respect to \( \mu_{\omega} \), we get:
\begin{align}
\frac{\partial}{\partial \mu_{\omega}} \left( - \int_0^T \lambda_{\omega}(t) \, dt \right) = - \int_0^T \frac{\partial \lambda_{\omega}(t)}{\partial \mu_{\omega}} \, dt.
\end{align}
By the simple calculation, \( \frac{\partial \lambda_{\omega}(t)}{\partial \mu_{\omega}} = 1 \), we arrive at:
\begin{align}
- \int_0^T 1 \, dt = -T.
\end{align}
Combining the two parts, we obtain the gradient:
\begin{align}
\frac{\partial \mathcal{L}}{\partial \mu_{\omega}} = \sum_{i=1}^{N_{\omega}} \frac{1}{\lambda_{\omega}(t_i^{\omega})} - T.
\end{align}
Thus, the gradient of the log-likelihood function \(\mathcal{L}\) with respect to \( \mu_{\omega} \) is given by:
\begin{align}
\frac{\partial \mathcal{L}}{\partial \mu_{\omega}} = \sum_{i=1}^{N_{\omega}} \frac{1}{\lambda_{\omega}(t_i^{\omega})} - T, 
\end{align}
which can be interpreted as follows: the first term \( \sum_{i=1}^{N_{\omega}} \frac{1}{\lambda_{\omega}(t_i^{\omega})} \) is the gradient based on the actual events, reflecting the relationship between the observed event times and the intensity function; the second term \( -T \) is the penalty term, which reflects the contribution of the expected number of events over the entire time window, ensuring that the model does not predict excessively high event frequencies during periods without events.

\subsection{B}\label{appendix:B}
The log-likelihood function for the high-dimensional Hawkes process involves two main terms: the log term and the integral term. The log-likelihood function for the parameter \( \alpha_{\omega, \omega'} \) is given by:

\begin{align}
\mathcal{L} = \sum_{i=1}^{N_{\omega}} \ln \left( \lambda_{\omega}(t_i^{\omega}) \right) - \int_0^T \lambda_{\omega}(t) \, dt,
\end{align}
where \( \lambda_{\omega}(t) \) is the intensity function for the process.

\textbf{Step 1: Log Term Gradient}

Recall that the log term is defined as:

\begin{align}
\sum_{i=1}^{N_{\omega}} \ln \left( \lambda_{\omega}(t_i^{\omega}) \right).
\end{align}

Taking the derivative with respect to \( \alpha_{\omega, \omega'} \), we deduce that:

\begin{align}
\frac{\partial}{\partial \alpha_{\omega, \omega'}} \ln \left( \lambda_{\omega}(t_i^{\omega}) \right) = \frac{1}{\lambda_{\omega}(t_i^{\omega})} \cdot \frac{\partial \lambda_{\omega}(t_i^{\omega})}{\partial \alpha_{\omega, \omega'}}.
\end{align}

We then substitute the intensity function into the above equation:

\begin{align}
\lambda_{\omega}(t) = \mu_{\omega} + \sum_{\omega' \in \Omega} \sum_{t_j^{\omega'} \leq t} \alpha_{\omega, \omega'} \phi_{\omega, \omega'}(t - t_j^{\omega'}), 
\end{align}

and calculate the derivative of \( \lambda_{\omega}(t) \) with respect to \( \alpha_{\omega, \omega'} \) as:

\begin{align}
\frac{\partial \lambda_{\omega}(t)}{\partial \alpha_{\omega, \omega'}} = \sum_{t_j^{\omega'} \leq t} \phi_{\omega, \omega'}(t - t_j^{\omega'}).
\end{align}

Thus, the gradient of the log term is:

\begin{align}
\frac{\partial}{\partial \alpha_{\omega, \omega'}} \ln \left( \lambda_{\omega}(t_i^{\omega}) \right) = \frac{1}{\lambda_{\omega}(t_i^{\omega})} \cdot \sum_{j=1}^{N_{\omega'}} e^{-\beta_{\omega,\omega'}(t_i^{\omega} - t_j^{\omega'})} \mathbf{1}_{\{t_j^{\omega'} < t_i^{\omega}\}}.
\end{align}

The full gradient for the log term is:

\begin{align}
\sum_{i=1}^{N_{\omega}} \frac{1}{\lambda_{\omega}(t_i^{\omega})} \sum_{j=1}^{N_{\omega'}} e^{-\beta_{\omega,\omega'}(t_i^{\omega} - t_j^{\omega'})} \mathbf{1}_{\{t_j^{\omega'} < t_i^{\omega}\}}.
\end{align}

\textbf{Step 2: Integral Term Gradient}

Recall that the integral term is given by:

\begin{align}
- \int_0^T \lambda_{\omega}(t) \, dt.
\end{align}
The derivative with respect to \( \alpha_{\omega, \omega'} \) is:

\begin{align}
\frac{\partial}{\partial \alpha_{\omega, \omega'}} \left( - \int_0^T \lambda_{\omega}(t) \, dt \right) = - \int_0^T \frac{\partial \lambda_{\omega}(t)}{\partial \alpha_{\omega, \omega'}} \, dt.
\end{align}

We calculate the derivative of of \(\lambda_{\omega}(t)\) with respect to \( \alpha_{\omega, \omega'} \) and get:

\begin{align}
\frac{\partial \lambda_{\omega}(t)}{\partial \alpha_{\omega, \omega'}} = \sum_{t_j^{\omega'} \leq t} e^{-\beta_{\omega,\omega'}(t - t_j^{\omega'})} \mathbf{1}_{\{t \geq t_j^{\omega'}\}}.
\end{align}

Therefore, the gradient of the integral term can be written as:

\begin{align}
- \int_0^T \sum_{t_j^{\omega'} \leq t} e^{-\beta_{\omega,\omega'}(t - t_j^{\omega'})} \mathbf{1}_{\{t \geq t_j^{\omega'}\}} \, dt.
\end{align}

We then replace the indicator function and perform the integral from \( t_j^{\omega'} \) to \( T \):

\begin{align}
\int_{t_j^{\omega'}}^T e^{-\beta_{\omega,\omega'}(t - t_j^{\omega'})} \, dt = \frac{1}{\beta_{\omega,\omega'}} \left( 1 - e^{-\beta_{\omega,\omega'}(T - t_j^{\omega'})} \right).
\end{align}

Thus, the gradient of the integral term is calculated as:

\begin{align}
- \sum_{j=1}^{N_{\omega'}} \frac{1}{\beta_{\omega, \omega'}} \left[ 1 - e^{-\beta_{\omega, \omega'}(T - t_j^{\omega'})} \right].
\end{align}

\textbf{Step 3: Final Gradient Expression}

Combining the log term and integral term gradients, we obtain the complete gradient with respect to \( \alpha_{\omega, \omega'} \) as:
\begin{align}
\frac{\partial \mathcal{L}}{\partial \alpha_{\omega, \omega'}} = 
\sum_{i=1}^{N_{\omega}} \frac{1}{\lambda_{\omega}(t_i^{\omega})} 
\sum_{j=1}^{N_{\omega'}} e^{-\beta_{\omega,\omega'}(t_i^{\omega} - t_j^{\omega'})} \mathbf{1}_{\{t_j^{\omega'} < t_i^{\omega}\}} 
-
\end{align}
\begin{align}
\sum_{j=1}^{N_{\omega'}} \frac{1}{\beta_{\omega,\omega'}} 
\left[ 1 - e^{-\beta_{\omega,\omega'}(T - t_j^{\omega'})} \right].
\end{align}

\subsection{C}\label{appendix:C}
The log-likelihood function for the high-dimensional Hawkes process involves two main terms: the log term and the integral term. The log-likelihood function for the parameter \( \alpha_{\omega, \omega'} \) is given by:

\begin{align}
\mathcal{L} = \sum_{i=1}^{N_{\omega}} \ln \left( \lambda_{\omega}(t_i^{\omega}) \right) - \int_0^T \lambda_{\omega}(t) \, dt,
\end{align}
where \( \lambda_{\omega}(t) \) is the intensity function for the process.

\textbf{Step 1: Log Term Gradient}

The log term is:

\begin{align}
\sum_{i=1}^{N_{\omega}} \ln \left( \lambda_{\omega}(t_i^{\omega}) \right).
\end{align}

Taking the derivative with respect to \( \alpha_{\omega, \omega'} \):

\begin{align}
\frac{\partial}{\partial \alpha_{\omega, \omega'}} \ln \left( \lambda_{\omega}(t_i^{\omega}) \right) = \frac{1}{\lambda_{\omega}(t_i^{\omega})} \cdot \frac{\partial \lambda_{\omega}(t_i^{\omega})}{\partial \alpha_{\omega, \omega'}}.
\end{align}

Using the intensity function:

\begin{align}
\lambda_{\omega}(t) = \mu_{\omega} + \sum_{\omega' \in \Omega} \sum_{t_j^{\omega'} \leq t} \alpha_{\omega, \omega'} \phi_{\omega, \omega'}(t - t_j^{\omega'}).
\end{align}

The derivative of \( \lambda_{\omega}(t) \) with respect to \( \alpha_{\omega, \omega'} \) is:

\begin{align}
\frac{\partial \lambda_{\omega}(t)}{\partial \alpha_{\omega, \omega'}} = \sum_{t_j^{\omega'} \leq t} \phi_{\omega, \omega'}(t - t_j^{\omega'}).
\end{align}

Thus, the gradient of the log term is:

\begin{align}
\frac{\partial}{\partial \alpha_{\omega, \omega'}} \ln \left( \lambda_{\omega}(t_i^{\omega}) \right) = \frac{1}{\lambda_{\omega}(t_i^{\omega})} \cdot \sum_{j=1}^{N_{\omega'}} e^{-\beta_{\omega,\omega'}(t_i^{\omega} - t_j^{\omega'})} \mathbf{1}_{\{t_j^{\omega'} < t_i^{\omega}\}}.
\end{align}

The full gradient for the log term is:

\begin{align}
\sum_{i=1}^{N_{\omega}} \frac{1}{\lambda_{\omega}(t_i^{\omega})} \sum_{j=1}^{N_{\omega'}} e^{-\beta_{\omega,\omega'}(t_i^{\omega} - t_j^{\omega'})} \mathbf{1}_{\{t_j^{\omega'} < t_i^{\omega}\}}.
\end{align}

\textbf{Step 2: Integral Term Gradient}

The integral term is:

\begin{align}
- \int_0^T \lambda_{\omega}(t) \, dt.
\end{align}

The derivative with respect to \( \alpha_{\omega, \omega'} \) is:

\begin{align}
\frac{\partial}{\partial \alpha_{\omega, \omega'}} \left( - \int_0^T \lambda_{\omega}(t) \, dt \right) = - \int_0^T \frac{\partial \lambda_{\omega}(t)}{\partial \alpha_{\omega, \omega'}} \, dt.
\end{align}

We know that:

\begin{align}
\frac{\partial \lambda_{\omega}(t)}{\partial \alpha_{\omega, \omega'}} = \sum_{t_j^{\omega'} \leq t} e^{-\beta_{\omega,\omega'}(t - t_j^{\omega'})} \mathbf{1}_{\{t \geq t_j^{\omega'}\}}.
\end{align}

Therefore, the gradient of the integral term is:

\begin{align}
- \int_0^T \sum_{t_j^{\omega'} \leq t} e^{-\beta_{\omega,\omega'}(t - t_j^{\omega'})} \mathbf{1}_{\{t \geq t_j^{\omega'}\}} \, dt.
\end{align}

We perform the integral from \( t_j^{\omega'} \) to \( T \):

\begin{align}
\int_{t_j^{\omega'}}^T e^{-\beta_{\omega,\omega'}(t - t_j^{\omega'})} \, dt = \frac{1}{\beta_{\omega,\omega'}} \left( 1 - e^{-\beta_{\omega,\omega'}(T - t_j^{\omega'})} \right).
\end{align}

Thus, the gradient of the integral term is:

\begin{align}
- \sum_{j=1}^{N_{\omega'}} \frac{1}{\beta_{\omega, \omega'}} \left[ 1 - e^{-\beta_{\omega,\omega'}(T - t_j^{\omega'})} \right].
\end{align}

\textbf{Final Gradient Expression}

Combining the log term and integral term gradients, the complete gradient with respect to \( \alpha_{\omega, \omega'} \) is:

\begin{align}
\frac{\partial \mathcal{L}}{\partial \alpha_{\omega, \omega'}} = 
\sum_{i=1}^{N_{\omega}} \frac{1}{\lambda_{\omega}(t_i^{\omega})} 
\sum_{j=1}^{N_{\omega'}} e^{-\beta_{\omega,\omega'}(t_i^{\omega} - t_j^{\omega'})} 
\mathbf{1}_{\{t_j^{\omega'} < t_i^{\omega}\}} 
- 
\end{align}
\begin{align}
\sum_{j=1}^{N_{\omega'}} \frac{1}{\beta_{\omega,\omega'}} 
\left[ 1 - e^{-\beta_{\omega,\omega'}(T - t_j^{\omega'})} \right].
\end{align}

\end{document}